\pdfoutput=1
\documentclass[11pt,a4paper]{article}
\usepackage{jcappub}

\usepackage[utf8]{inputenc}
\usepackage{graphicx}
\usepackage{amsmath,amssymb}   
\usepackage[normalem]{ulem}
\usepackage{mathtools}
\usepackage[titletoc]{appendix}
\usepackage{datetime}
\usepackage{upgreek}
\usepackage{rotating}
\usepackage{makecell}
\usepackage{verbatim}
\usepackage{array}
\usepackage{afterpage}
\usepackage{hyperref}

\def\ha{{\mathcal{H}}}

\def\idn{{\Gamma_{\mathrm{DM}-\upnu}}}
\def\ind{{\Gamma_{\upnu-\mathrm{DM}}}}
\def\acrit{{a_\mathrm{max}}}
\def\unudm{{u_{\upnu\mathrm{DM}}}}
\def\unudmO{{u_{\upnu\mathrm{DM},0}}}
\def\nnudm{{n_{\upnu\mathrm{DM}}}}

\def\dm{{\mathrm{DM}}}
\def\dmu{{\dot{\kappa}_{\upnu\dm}}}
\def\lm{{l_\mathrm{max}}}
\def\ur{{\mathrm{ur}}}
\def\anudec{{a_{\upnu,\mathrm{dec}}}}
\def\mp{{M_\mathrm{P}}}
\def\ufa{{\mathrm{UFA}}}

\begin{document}

\title{Comprehensive Study of Neutrino-Dark Matter Mixed Damping}

\author[a,1]{Julia Stadler
\note{\hyperlink{https://orcid.org/0000-0001-5888-023X}{https://orcid.org/0000-0001-5888-023X}}}

\author[b,c,d,2]{C{\'e}line B{\oe}hm 
\note{\hyperlink{https://orcid.org/0000-0002-5074-9998}{https://orcid.org/0000-0002-5074-9998}}}

\author[e]{Olga Mena
\note{\hyperlink{https://orcid.org/0000-0001-5225-975X}{https://orcid.org/0000-0001-5225-975X}}}

\affiliation[a]{Institute for Particle Physics Phenomenology, Durham University, South Road, Durham, DH1 3LE, United Kingdom}
\affiliation[b]{School  of  Physics,  Physics  Road,  The  University  of  Sydney,  NSW  2006 Camperdown,  Sydney,  Australia}
\affiliation[c]{LAPTH, U. de Savoie, CNRS, BP 110, 74941 Annecy-Le-Vieux, France}
\affiliation[d]{Perimeter institute, 31 Caroline Street N., Waterloo, Ontario,  N2L 2Y5, Canada}
\affiliation[e]{Instituto de F{\'i}sica Corpuscular (IFIC), CSIC-Universitat de Val{\`e}ncia,\\
Apartado de Correos 22085,  E-46071, Spain}

\emailAdd{julia.j.stadler@durham.ac.uk}
\emailAdd{celine.boehm@sydney.edu.au}
\emailAdd{omena@ific.uv.es}

\abstract{
Mixed damping is a physical effect that occurs when a heavy species is coupled to a relativistic fluid which is itself free streaming. As a cross-case  between collisional damping and free-streaming, it is crucial in the context of neutrino-dark matter interactions. In this work, we establish the parameter space relevant for mixed damping, and we derive an analytical approximation for the evolution of dark matter perturbations in the mixed damping regime to illustrate the physical processes responsible for the suppression of cosmological perturbations. Although extended Boltzmann codes implementing neutrino-dark matter scattering terms automatically include mixed damping, this effect has not been systematically studied. In order to obtain reliable numerical results, it is mandatory to reconsider several aspects of neutrino-dark matter interactions, such as the initial conditions, the ultra-relativistic fluid approximation and high order multiple moments in the neutrino distribution. Such a precise treatment ensures the correct assessment of the relevance of mixed damping in neutrino-dark matter interactions.
}

\maketitle

\section{Introduction}
Dark matter (DM) is a required ingredient in our universe to  explain e.g. the  galactic rotation curves, gravitational lensing, Cosmic Microwave Background (CMB) measurements and the growth of matter perturbations. Large-scale-structures observations by galaxy surveys and other measurements seem to be perfectly consistent with a Cold Dark Matter (CDM) component, that is, with a pressureless, non-interacting fluid. However, small-scale discrepancies \cite{Bullock:2017xww} indicate that the properties of dark matter could be more complex than the CDM hypothesis.

In this regard, dark matter interactions with Standard Model particles can change observations at small scales \cite{Boehm:2000gq,Boehm:2001hm,Chen:2002yh,Boehm:2003xr, Boehm:2004th,Boehm:2014vja,Mangano:2006mp,Serra:2009uu,Wilkinson:2013kia,Schneider:2014rda,Schewtschenko:2015rno,Escudero:2015yka,Schewtschenko:2014fca,Ali-Haimoud:2015pwa,Murgia:2017lwo,Wilkinson:2016hdl,Diacoumis:2017hff,DiValentino:2017oaw,Campo:2017nwh,Diacoumis:2018ezi,Escudero:2018thh,Kumar:2018yhh,Stadler:2018jin,Stadler:2018dsa,Lopez-Honorez:2018ipk,Bringmann:2013vra,Aarssen:2012fx}. Mixed damping \cite{Boehm:2000gq,Boehm:2003xr, Boehm:2004th} refers to the physical damping phenomenon in which dark matter is kinetically coupled to another species which itself is free streaming. In the context of interacting dark matter, mixed damping is particularly relevant for dark matter-neutrino interactions, the main focus of this study, but it might also apply to other dark matter interactions, e.g. with a dark radiation component \cite{Kamada:2017oxi}. Note that a similar effect, Silk damping \cite{Silk:1967kq}, exists in the evolution of the baryon-photon plasma, where photons decouple briefly before the end of the baryon drag epoch \cite{Hu:1994uz,Hu:1995en}. However, in the interacting dark matter scenarios considered in this study the dark matter decoupling history is very distinct from the baryonic one. Dark matter perturbations evolve in the mixed damping regime for a much longer time, while simultaneously radiation perturbations are largely unaffected. Hence, the effect of mixed damping on dark matter and neutrino perturbations considerably differs from what is observed in the baryon-photon sector.

The mixed damping effect is a cross between collisional damping and free-streaming; dark matter perturbations are damped because the dark matter follows the free-streaming neutrinos. The condition for mixed damping to occur is thus
\begin{equation}
\idn > H > (\Gamma_\upnu \equiv\Gamma_{\upnu-\mathrm{e}} + \ind)  \, ,
\label{eq: mixed-damping-condition-1}
\end{equation}

\noindent where the collision rates between dark matter and neutrinos are  given by
\begin{subequations}
\begin{align}
\ind &= n_\dm\, \sigma_{\upnu\dm} 
\label{eq: param-rate-nudm}\,,\\
\idn &= \frac{4\rho_\upnu}{3\rho_\dm}\, \ind\,.
\label{eq: param-rate-dmnu}
\end{align}
\label{eq: param-rate-both}
\end{subequations}
The condition in Eq.~(\ref{eq: mixed-damping-condition-1}) assumes that dark matter scattering is the only non-standard interaction of the neutrino species beyond the usual weak interactions with electrons occurring at a rate $\Gamma_{\upnu-\mathrm{e}} = \sigma_{\upnu-\mathrm{e}}\, n_\mathrm{e}$.

To progress further, one needs to specify the dark matter number density $n_\dm$. The observed dark matter relic density suggests that, if the dark matter was produced thermally, it has either annihilated or decayed since then~\footnote{There may be other mechanism in the thermal scenario that could explain why the dark matter  number density nowadays is so small. Alternatively dark matter may have been produced non-thermally.}. In a thermal annihilating dark matter scenario, where the dark matter and anti dark matter number densities are supposed to be exactly the same, the observed relic density imposes a condition on the ratio of the dark matter mass to the freeze-out temperature (for a given annihilation cross section), which can be used to determine whether one needs to account for the evolution of the dark matter number density while neutrinos are decoupling~\cite{Boehm:2004th}. In the following, we will disregard such a scenario and assume that dark matter has already annihilated when the neutrino fluid decouples. As a result, the bounds that we quote in the following should be used with precaution when dark matter is lighter than $m_{\dm}  \lesssim O(\rm{MeVs})$.

In the past, several constraints on dark matter-neutrino interactions have been derived from a plethora of cosmological observations. Forecasts for next-generation of CMB and large scale structure surveys are also available in the literature, see e.g.~\cite{Escudero:2015yka}. Constraints are given in terms of a single parameter, namely 
\begin{equation}
\unudm = \frac{\sigma_{\upnu\dm}}{\sigma_\mathrm{Th}}\,\left(\frac{m_\dm}{100\,\mathrm{GeV}}\right)^{-1}\,,
\label{eq: unudm}
\end{equation}
where the dark matter-neutrino scattering cross section is normalised to the Thomson scattering rate $\sigma_\mathrm{Th}$. The dark matter-neutrino scattering cross-section may be a power-law of the neutrino temperature, i.e. $\sigma_{\upnu\dm}\propto T_\upnu^{\,\nnudm}$. It is convenient to further define
\begin{equation}
\unudm = \unudmO \times a^{-\nnudm}\,,
\label{eq: nnudm}
\end{equation}
where $\unudmO$ is a constant. The most recent analyses provide $\unudm \le (4.5-9.0)\times 10^{-5}$ for $\nnudm=0$ and  $\unudmO \le (3.0-5.4) \times 10^{-14}$ for $\nnudm=2$ \cite{Wilkinson:2014ksa,Escudero:2015yka,DiValentino:2017oaw}. Mixed damping is an important effect for these values, as we discuss in Sec.~\ref{sec: parameter-dm}. Although the effects of mixed damping are automatically accounted for in the Boltzmann hierarchy for collisional neutrinos, a systematic description of the mixed damping regime is lacking to date. Furthermore, there is a number of subtleties which have to be taken into account in the numerical evolution of perturbations if dark matter interacts with neutrinos. In the present manuscript we focus on all these aspects.

The structure of this paper is as follows. We start by examining the parameter space for which mixed damping is important in Sec.~\ref{sec: parameter}, describing the conditions which Eq.~(\ref{eq: mixed-damping-condition-1}) imposes on the scale factor at decoupling, the scales which can be affected by mixed damping and the values of the interactions strength parameter $\unudm$. We then turn to the numerical evolution of perturbations in Sec.~\ref{sec: numerical}, where we introduce the Boltzmann equations for neutrino-dark matter scattering. We adopt a self consistent approach to account for the angular dependence of the matrix elements for the scattering process in the evolution of higher-order multipoles.  We shall also examine the validity of the ultra-relativistic fluid approximation in the presence of dark matter-neutrino interactions and present the required modifications to the initial conditions in the presence of a coupled fluid. In order to investigate the mixed damping physics we derive an analytical approximation for the evolution of dark matter perturbations in this scenario in Sec.~\ref{sec: mixed-damping-effect}. We conclude in Sec.~\ref{sec: conclusions}.

\section{Mixed damping regime and parameter space}
\label{sec: parameter}

As mentioned in the introduction, mixed damping occurs when neutrinos decouple from dark matter before the reverse happens. In general, the two interaction rates $\Gamma_{\dm-\upnu}$ and $\Gamma_{\upnu-\dm}$  are not expected to be equal because of the differences between the neutrino and dark matter number densities (c.f. Eq.~(\ref{eq: param-rate-both})).  In the Standard Model, neutrinos kinetically decouple from the thermal bath when they decouple from the electrons. However, in models where neutrinos interact with dark matter, the neutrino  interaction rate is given by $\Gamma_{\upnu} \equiv \Gamma_{\upnu-\dm} + \Gamma_{\upnu-\mathrm{e}}$. The neutrino kinetic decoupling is thus determined by their last interactions, which might be with either the electrons (if $\Gamma_\upnu \sim \Gamma_{\upnu-\mathrm{e}}\simeq H$) or dark matter (if $\Gamma_{\upnu} \sim \Gamma_{\upnu-\dm}  \simeq H$). Each option corresponds to a different dark matter decoupling epoch and to a different maximum damping scale, which we further examine in Sec.~\ref{sec: parameter-waek} and Sec.~\ref{sec: parameter-dm}, respectively.

In any case, however, mixed damping can only occur if there is an epoch where $\Gamma_{\dm-\nu} > H > \Gamma_{\nu}$. This condition provides a lower bound on the scales which can be affected by mixed damping and a corresponding minimum scattering rate. It is set by the simultaneous decoupling of dark matter from neutrinos and of neutrinos from electrons and leads to  $u_{\upnu\dm,0}^\mathrm{min}> 2.4\times 10^{-14}$, $1.4\times 10^{-33}$ and $7.4\times 10^{-53}$ for $\nnudm=0$, $2$ and $4$ respectively. The scale factor at decoupling in this configuration is $a_\mathrm{min} =  a\left(T_\upgamma=1\,\mathrm{MeV}\right) = 2.35\times 10^{-10}$ and the corresponding "mixed damping scale" is rather small, lying beyond the current experimental reach: $r_\mathrm{min} = \left(a\,H\right)^{-1}_{T_\upgamma=1\,\mathrm{MeV}} = 0.11$~kpc, or $M_\mathrm{min} = 0.2 M_\odot $.

\subsection{Neutrino decoupling by weak interactions}
\label{sec: parameter-waek}

The mixed damping effect is the sole process responsible for erasing dark matter perturbations larger than $r_\mathrm{min}$ when the neutrino decoupling is set by its Standard Model interactions, i.e. $\Gamma_{\dm-\upnu} > \Gamma_{\upnu-\mathrm{e}} > \Gamma_{\upnu-\dm}$ (assuming a monotonous thermal evolution of both dark matter and neutrinos in the early Universe). In this case, the magnitude of the effect is determined by the dark matter decoupling only ($\Gamma_{\dm-\nu} \simeq H$); the later the decoupling, the bigger the effect.  Still, $\Gamma_{\dm-\upnu}$ can not become arbitrarily large. Indeed the condition $\Gamma_{\upnu-\dm} < \Gamma_{\upnu-\mathrm{e}}$ in combination with Eq.~(\ref{eq: param-rate-dmnu}) implies that the condition for dark matter decoupling, $\Gamma_{\dm-\upnu} \simeq H$ can be recasted as
\begin{equation}
\frac{4\rho_\upnu}{3\rho_\dm} \ \Gamma_{\upnu-\mathrm{e}} \simeq H\,,
\end{equation}
for the maximum neutrino-dark matter interaction rate value allowed. The corresponding maximum scale on which mixed damping can occur, $r_\mathrm{wdec}$, is $79\,\mathrm{kpc}$, $2.6\,\mathrm{kpc}$ and $0.8\,\mathrm{kpc}$ for $\nnudm=0, 2$ and $4$ respectively (or, in terms of enclosed mass, $M_\mathrm{wdec} = 8\times 10^7\,M_\odot \,,~ 3\times 10^3\,M_\odot\,,~ 1\times 10^2\,M_\odot$ for the respective power laws).

Neutrino decoupling by weak interactions can only occur if the dark matter interaction rate lies below a maximum value. Combined with the previous section's discussion, the mixed damping condition on the interaction strength parameter in this regime is $u_{\upnu\dm,0}^ \mathrm{min} < \unudmO <u_{\upnu\dm,0}^\mathrm{wdec}$. In numbers, the upper bound is $u_{\upnu\dm,0}^\mathrm{wdec} = 1.3\times 10^{-8}\,,~ 7.4\times10^{-28}\,,~ 4.1\times 10^{-47}$ for $\nnudm=0, 2$ and $4$ respectively. For these limiting parameters we depict the evolution of the interaction rates in Fig.~\ref{fig: parameterspace-scattering-rates}, where the left edge of the green shaded region indicates the time of neutrino decoupling in the Standard Model.

\subsection{Neutrino decoupling set by dark matter interactions} 
\label{sec: parameter-dm}

Mixed damping also plays an important role if the neutrino kinetic decoupling is determined by the neutrino-dark matter interactions. This is relevant for larger modes, which are easier to access  observationally. To illustrate the situation, first consider a small mode, which enters the horizon during a period where $\Gamma_{\dm-\upnu} > \Gamma_{\upnu-\dm} > H$. This mode will be initially  subject to collisional damping and, upon neutrino decoupling, will experience a transition to the mixed damping regime. A larger mode, on the other hand, enters the horizon later and thus can be subject to mixed damping only. However, the transition from the collisional to the mixed damping regime can only occur if the ratio of densities in Eq.~(\ref{eq: param-rate-dmnu}) is larger than unity by the time dark matter decouples from the neutrino fluid. The corresponding upper limit on the decoupling time is
\begin{equation}
\left.\frac{4\,\rho_\upnu}{3\,\rho_\dm}\right|_{\acrit} = 1
\quad \Leftrightarrow \quad 
\acrit = 1.9\times10^{-4}\left(\frac{\Omega_\dm\, h^2}{0.1186}\right)^{-1}.
\end{equation}

The largest scales that will be affected by mixed damping have therefore to enter the horizon before $\acrit$. The comoving Hubble radius at $\acrit$ is 
\begin{equation}
r_\mathrm{max}  
= \frac{71.0\,\mathrm{Mpc}\,\left(\frac{0.1186}{\Omega_\dm\, h^2}\right)}
{\sqrt{1.0 + 0.069\left(\frac{\Omega_\mathrm{b}\, h^2}{0.0223}\right)\left(\frac{0.1186}{\Omega_\dm\, h^2}\right)}}\,,
\end{equation}
which roughly corresponds to a mass of $M_\mathrm{max} \simeq  6\times 10^{16}\,M_\odot$.

Finally, the criterion of neutrinos decoupling from the dark matter before $\acrit$ translates into a maximum value of the scattering rate
\begin{align}
u_{\upnu\dm,0}^\mathrm{max} &= 1.97\times 10^{-2}\times a_\mathrm{max}^\nnudm\times\left(\frac{0.1186}{\Omega_\dm\, h^2}\right)^{2}\,\sqrt{1.0 + 0.066\left(\frac{\Omega_\mathrm{b}\, h^2}{0.0223}\right)\left(\frac{0.1186}{\Omega_\dm\, h^2}\right)}\,.
\end{align}
We find $u_{\upnu\dm,0}^\mathrm{crit} \simeq 1.98 \times 10^{-2}$, $7.12 \times 10^{-10}$ and
$2.56 \times 10^{-17}$, for $\nnudm = 0$, $2$ and $4$ respectively.
Note that these values fall within the current observational limits~\cite{Wilkinson:2016hdl,Escudero:2015yka}, indicating the importance of mixed damping while deriving bounds on scenarios with dark matter-neutrino interactions. 

\begin{figure}
    \centering
    \includegraphics{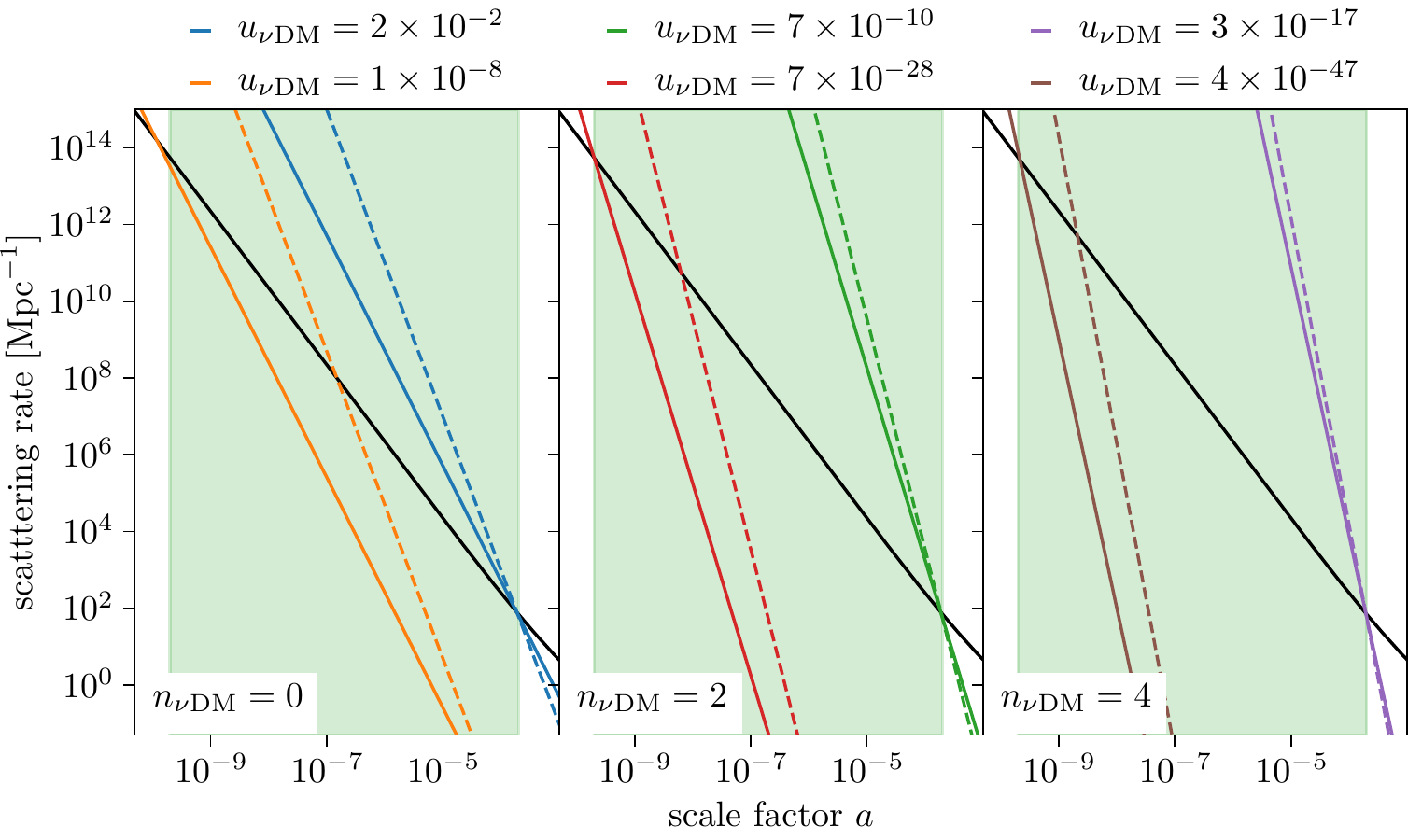}
    \caption{The evolution of $\Gamma_{\nu-\dm}$ (solid lines), $\Gamma_{\dm-\nu}$ (dashed lines) and the Hubble rate $H$ (black solid line) as a function of the scale factor $a$ for three possible temperature dependent scenarios and two possible values of the normalised interacting cross section. The green region indicates the dark matter decoupling times at which mixed damping is possible.}
    \label{fig: parameterspace-scattering-rates}
\end{figure}

The evolution of scattering rates in this limiting case is depicted in Fig.~\ref{fig: parameterspace-scattering-rates} along with the Hubble rate. The right edge of the green region indicates $a_\mathrm{max}$. Irrespective of the value chosen for $\unudmO$ and $\nnudm$ the dark matter and the neutrino scattering rate are always the same at $\acrit$. Mixed damping occurs whenever $u_{\nu\dm,0}^\mathrm{min} < \unudmO <u_{\nu\dm,0}^\mathrm{max}$, or, in terms of Fig.~\ref{fig: parameterspace-scattering-rates}, whenever the neutrino and dark matter decoupling times fall within the green region. Depending on the size of a mode it might either be subject to mixed damping solely or to a period of collisional damping followed by mixed damping.

\section{Numerical evolution of cosmological perturbations}
\label{sec: numerical}

Up to now we have discussed the mixed damping on the level of the homogeneous background evolution. In order to investigate in detail how mixed damping affects cosmological perturbations, we now turn to the linearised set of Einstein and Boltzmann equations. After showing how the relevant equations are altered by the presence of a scattering term between dark matter and neutrinos, we shall address several numerical subtleties absolutely mandatory to ensure the correctness of the formalism.

\subsection{Boltzmann Equations}
\label{sec: boltzmann-equations}

In the Newtonian gauge,
\begin{equation}
ds^2 = a^2(\tau)\left[-\left(1+2\psi\right)d\tau^2 + \left(1-2\phi\right)dx_i dx^i\right]   \,, 
\end{equation}
where $\phi$ and $\psi$ are the metric perturbations, the scattering processes between dark matter and massless neutrinos are governed, for the dark matter fluid, by its evolution equations,
\begin{subequations}
\begin{align}
\dot{\delta}_\dm &= -\theta_\dm + 3\dot{\phi}\,,\\[5pt]
\dot{\theta}_\dm &= k^2\psi -\ha\,\theta_\dm - R\,\dmu\left(\theta_\dm - \theta_\upnu\right)\,.
\end{align}
\label{eq: dm-evolution}
\end{subequations}
Here $R = 4\rho_\upnu/(3\rho_\dm)$, the scattering rate is defined as $\dmu=a\, n_\dm\, \sigma_{\nu\dm}$ and the reduced Hubble rate given by $\ha = a\,H$. For massless neutrinos the modified Boltzmann hierarchy is~\cite{Ma:1995ey,Cyr-Racine:2015ihg}
\begin{subequations}
\begin{align}
\dot{\delta}_\upnu &= -\frac{4}{3}\theta_\upnu + 4\dot{\phi}\,,\\[5pt]
\dot{\theta}_\upnu &= k^2\left(\frac{\delta_\upnu}{4} - \sigma\upnu\right) + k^2\psi - \dmu\left(\theta_\upnu - \theta_\dm\right)\,,\\[5pt]
2\dot{\sigma}_\upnu &= \frac{8}{15}\theta_\upnu - \frac{3}{5}k F_{\upnu,3} - \alpha_2\,\dmu\sigma\upnu
\label{eq: boltzmann-nu-shear}\,,\\[5pt]
\dot{F}_{\upnu,l} &= \frac{k}{2l+1}\left[l F_{\upnu,l-1} - (l+1)F_{\upnu,l+1}\right] - \alpha_l\,\dmu F_{\upnu,l}\,,\\[5pt]
\dot{F}_{\upnu,\lm} &= k\left[F_{\upnu,\lm-1} - \frac{\lm+1}{k\tau}F_{\upnu,\lm}\right] - \alpha_l\,\dmu F_{\upnu,\lm}\,.
\end{align}
\label{eq: boltzmann-nu}
\end{subequations}
The angular coefficients $\alpha_l$, which appear in the interaction terms of the higher-order multipoles, are generally of $\mathcal{O}(1)$. Nevertheless, their precise numerical value is set by the dependence of the matrix element for the scattering process $|\mathcal{M}_{\nu\dm}|^2$ on the cosine of the angle between the incoming and the outgoing neutrino $\mu$. Previous works on dark matter-neutrino scattering have adopted different choices for $\alpha_l$, concretely $\alpha_2=2$ and $\alpha_l = 1$ for $l\ge3$~\cite{Escudero:2015yka}, or $\alpha_2 = 9/5$ and $\alpha_l = 1$ for $l\ge3$~\cite{DiValentino:2017oaw}. Here, we follow Ref.~\cite{Cyr-Racine:2015ihg}, which provides a self-consistent formalism to compute the higher-order multipole coefficients,
\begin{equation}
\alpha_l = \frac
{\int dp \ p^4 \left(\frac{\partial f\upnu}{\partial p}\right)\left[A_0(p) - A_l(p)\right]}
{\int dp \ p^4 \left(\frac{\partial f\upnu}{\partial p}\right)\left[A_0(p) - A_1(p)\right]}\,.
\end{equation}
Neutrinos follow a Fermi-Dirac statistic, denoted by $f_\upnu$, and the matrix element affects the coefficients via its projection on the Legendre polynomials $P_l$
\begin{equation}
A_l (p) = \frac{1}{2} \int_{-1}^{1} d\mu \  P_l(\mu)\left.\left(\frac{1}{\eta_\dm\,\eta_\upnu}\left\vert\mathcal{M}_{\nu\dm}\right\vert^2\right) \right\vert_{\substack{t=2p^2(\mu-1)\hfill\\ s=m_\dm^2+2 \,  m_\dm \, p}}\,.
\end{equation}
In the low energy limit, the momentum $p$ of the neutrino does not to change in the scattering process, $t$ and $s$ are Mandelstam variables, and $\eta_\dm$ and $\eta\upnu$ denote the internal degrees of freedom of the dark matter particle and neutrinos respectively. Within this formalism, the scattering rate can be expressed as~\cite{Cyr-Racine:2015ihg}
\begin{equation}
\dmu = \frac{a\,n_\dm}{128\,\uppi^3\,m^2_\dm}\,\frac{\eta_\upnu}{\rho_\upnu}\,\int^\infty_0 dp \left(\frac{\partial f_\upnu}{\partial p}\right) p^4 \left[A_0(p) - A_1(p)\right]\,.
\end{equation}
We make use of  the classification of neutrino-dark matter interaction scenarios introduced in \cite{Campo:2017nwh} to give an overview of the derived values for $\dmu$ and $\alpha_l$ in Tab.~\ref{tab: interaction-scenarios} in Appendix~\ref{sec:secA}. To evolve cosmological perturbations numerically, we have modified the publicly available Boltzmann solver code CLASS
\footnote{Our modified CLASS version is publicly available and can be downloaded from \hyperlink{https://gitlab.dur.scotgrid.ac.uk/dm-interactions/class_v2.7_ndm.git}{https://gitlab.dur.scotgrid.ac.uk/dm-interactions/class\_v2.7\_ndm.git}} (version v2.7), introducing three new input parameters. Namely, $\unudm$ represents the coupling strength parameter (see Eq.~(\ref{eq: unudm})), $n_{\nu\dm}$ governs the temperature dependence of the cross-section (see Eq.~(\ref{eq: nnudm})) and $\alpha_l$ refers to the higher-order multipole coefficients appearing in Eq.~(\ref{eq: boltzmann-nu}). 

\subsection{The Ultra-relativistic Fluid Approximation}
\label{sec: ufa}
The evolution of neutrino perturbations is described by an infinite hierarchy of moment equations (c.f. Eqs.~(\ref{eq: boltzmann-nu})), that must be truncated at some maximum multipole $\lm$. At early times $\lm$ is typically of $\mathcal{O}(10)$. Once neutrino perturbations are well inside the horizon, neutrino multipoles over the range $2 < l \ll k\tau$ are suppressed, and the evolution of the highest and the lowest multipoles decouples \cite{Blas:2011rf}. Note that, during the radiation dominated era, the size of the comoving Hubble radius is given by the conformal time, $\ha^{-1} \simeq \tau$. In this regime, CLASS employs the ultra-relativistic fluid approximation (UFA) \cite{Blas:2011rf}, which truncates the multipole hierarchy after $l=2$ once that $k\tau > \left(k\tau\right)_{\mathrm{UFA}}$.
The advantage is twofold. First, the computational costs to describe the late time evolution of neutrino perturbations are reduced. Second, to avoid unphysical reflections, which are caused by the inevitable truncation of the a priori infinite Boltzmann hierarchy, for a mode of wavenumber $k$ at some time $\tau$ one has to choose the maximum multipole moment $\lm\ge k\tau$. Truncating the Boltzmann hierarchy in a consistent way as earlier as possible then allows to choose a smaller value for $\lm$ during the early evolution and hence also benefits the computational costs prior to the UFA \footnote{The default values for these two parameters in CLASS are $\left(k\tau\right)_{\mathrm{UFA}}=30$ and $\lm=17$.}. In neutrino-dark matter interacting schemes the UFA method can also be applied, but the conditions must be carefully revised. This is the aim of this section.

The UFA truncation scheme in general differs from the ordinary truncation scheme for Boltzmann equations proposed in Ref.~\cite{Ma:1995ey} and assumes free streaming neutrinos. In the $\Lambda$CDM scenario this  assumption is always valid on all scales larger than $r_\mathrm{min}$, which enter the horizon after neutrino-electron decoupling. But, as we discussion in Sec.~\ref{sec: parameter-dm}, the coupling between dark matter and neutrinos can delay the epoch of neutrino free streaming. Previous studies dealing with neutrino-dark matter interactions have tried to generalise the existing UFA expression, just by including an interaction term. Since the origin of such a term is not fully understood, we follow here a different avenue. We choose $\left(k\tau\right)_{\rm{UFA}}$ large enough to make sure that neutrinos have decoupled from dark matter when the UFA starts. Then, we evolve the neutrino perturbations accordingly to Eqs.~(\ref{eq: boltzmann-nu}) while the coupling to dark matter is active. Once neutrinos have decoupled from dark matter, we switch to the standard UFA truncation formula which does not include interactions, c.f. Eq.~(\ref{ufa: truncation-equation}). In parallel, we ensure that $\lm$ is large enough to avoid unphysical reflections while the Boltzmann hierarchy is evolved.

The conditions on $\left(k\tau\right)_\mathrm{UFA}$ are detailed in Appendix~\ref{sec: ufa-appendix}, where we also analyse how the premature application of the ultra-relativistic fluid approximation affects the cosmological observables. Since the effect of the UFA treatment on CMB temperature and polarisation spectra is well below the experimental sensitivity for $\unudm=4.5\times 10^{-5}$ ($\nnudm=0$) and for $\unudmO=5.4\times10^{-14}$ ($\nnudm=2$), the constraints obtained in Refs.~\cite{DiValentino:2017oaw,Wilkinson:2014ksa} remain valid, regardless the UFA was the default one.

Discrepancies are however evident in the matter power spectrum, c.f. Fig.~\ref{fig: ufa-impact-pk}. For those interaction strengths interesting to mixed damping and to cosmological constraints, the differences only arise at small scales on which the matter power spectrum is already suppressed by several orders of magnitude with respect to the $\Lambda$CDM prediction. Differences at this level will be completely erased by non-linear effects and, more importantly, are negligible when compared to the expected systematic effects in galaxy surveys. In Appendix~\ref{sec: ufa-appendix} we further illustrate this by estimating the effect of the UFA treatment on the predicted number of Milky Way satellites. Discrepancies arising from the treatment of the UFA regime are well below Poisson scatter, which the counting is subject to.

\subsection{Initial conditions}
\label{sec: initial-conditions}
If neutrinos are coupled to dark matter this affects not only the evolution of perturbations but also their initial conditions. The reason for this is obvious from the traceless space-space component of the linearised Einstein equations
\begin{equation}
k^2\left(\phi - \psi\right) = 12\uppi\, G\, a^2 \left(\rho + p\right)\sigma\,.
\label{eq: einstein}
\end{equation}
In the radiation dominated era the anisotropic stress is dominated by neutrino perturbations, and $\sigma_\upnu$ will be suppressed if neutrinos scatter off dark matter. Even though we aim to study the mixed damping regime, where neutrinos are free streaming, we have shown in Sec.~\ref{sec: parameter} that the transition from the collisional to the mixed damping regime is continuous and smooth. A perturbation of fixed wavelength can start its evolution in the collisional regime and then evolve towards the mixed damping one.

In Appendix~\ref{sec: ini-appendix} we give the initial conditions in the tight neutrino-dark matter coupling limit. Obviously, these expressions are only applicable to the smallest modes, which enter the horizon earliest. The largest modes, on the other hand, enter the horizon when neutrinos have completely decoupled and hence are described by the default initial conditions. The ratio $\Gamma_{\upnu-\dm}/H$ for each mode prior at horizon entry determines which case should one follow. Namely, if $\Gamma_{\upnu-\dm}/H \ll 1$, we proceed with the default initial conditions. In the opposite case, we start the integration sufficiently early such that $\Gamma_{\upnu-\dm}/H \gg 1$ and use instead the tightly coupled initial conditions. Nevertheless, by comparing  the CMB and the matter power spectra obtained for the default initial conditions to those merging from our devoted analyses, we find that the differences in the CMB spectra are below the $\upmu\mathrm{K}^2$ level and that the relative difference in the matter power spectrum reaches $10\%$ only in that range where this observable is already suppressed by several orders of magnitude. Hence the constraints and conclusions obtained in previous works on the phenomenology of neutrino-dark matter scattering (c.f. Refs~\cite{Escudero:2015yka,Wilkinson:2014ksa,DiValentino:2017oaw}) are stable against the initial conditions.

\section{The physics of mixed damping}
\label{sec: mixed-damping-effect}

Having discussed all the relevant technical aspects required to obtain accurate predictions for the evolution of perturbations in the mixed damping scenario, we now turn to the evolution of individual modes and discuss how these are affected by mixed damping. We start with a brief discussion of the evolution of perturbations in the canonical $\Lambda$CDM case and compare how the decoupling history of the baryon photon plasma differs from neutrino-dark matter decoupling in the mixed damping regime. These differences are important for the derivation of an analytical approximation to the evolution of dark matter perturbations presented in the following. Finally, we compare our analytical results to the full numerical results to elucidate the physical mechanism behind the mixed damping effect.

The scale factors at which dark matter and neutrinos decouple from the respective other species relate as
\begin{equation}
a_{\upnu,\mathrm{dec}} = \left(\frac{a_{\dm,\mathrm{dec}}}{\acrit}\right)^{\frac{1}{n_{\upnu\dm}+1}}\,a_{\dm,\mathrm{dec}}\,.
\label{eq: nu-decoupling-n-dependence}
\end{equation}
Mixed damping requires that $a_{\dm,\mathrm{dec}} < \acrit$. For a fixed dark matter decoupling time, neutrinos decouple earlier the smaller the power law index $\nnudm$ is, and the mixed damping regime lasts longer, see Fig.~\ref{fig: parameterspace-scattering-rates}. Correspondingly, a larger range of modes is subject to mixed damping only, making this regime more accessible. In addition, the numerical solution is more stable for small values of $\nnudm$, because for a given decoupling time the scattering rates in Eqs.~(\ref{eq: boltzmann-nu}) and (\ref{eq: dm-evolution}) are smaller when the numerical integration starts, and hence the system of differential equations is less stiff. We therefore focus the discussion here on the $\nnudm=0$ case. The generalisation to other temperature dependencies of the scattering cross section is however straightforward.

\subsection{Comparison of decoupling histories}

\begin{figure}
\centering
\includegraphics[]{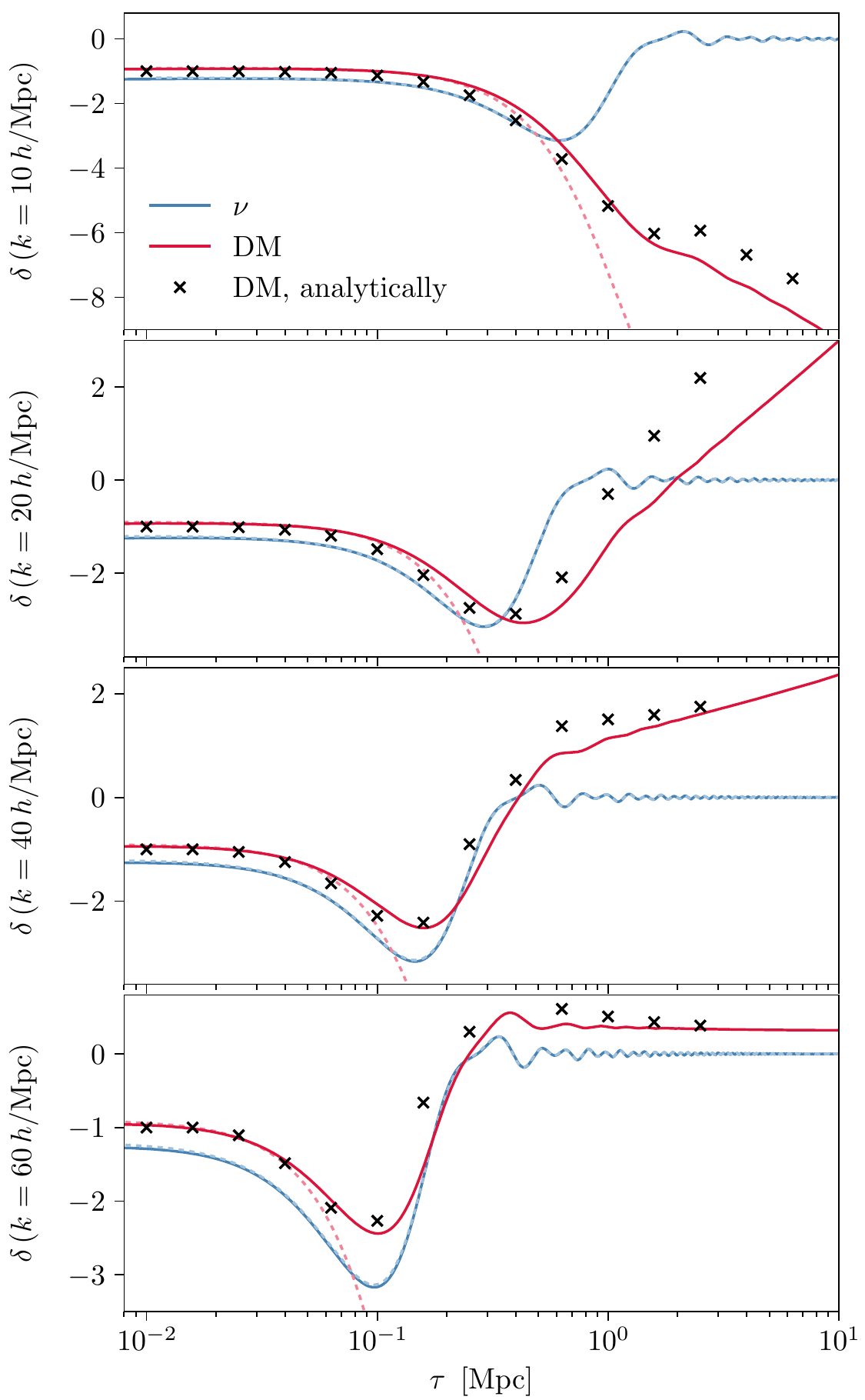}
\caption{
Evolution of density perturbations for individual modes in the $\Lambda$CDM case (dashed, pastel lines) and for a neutrino-dark matter interaction strength of $\unudm=10^{-6}$ and $\nnudm=0$ (dark, solid lines). The analytic prediction for the dark matter  evolution obtained from the integration of Eq.~(\ref{eq: dm-evolution}) is depicted by the black points.}
\label{fig: mixed-damping-mode}
\end{figure}

Among all species, the evolution of dark matter perturbations in the $\Lambda$CDM scenario is probably the simplest: they are constant while the Hubble radius is smaller than the mode's size and receive a boost upon horizon crossing which sets them in a growing mode. The growth is proportional to $\log a$ while the universe is radiation dominated and proceeds as $\delta_\dm \propto  a$ once the universe has transitioned to matter domination \cite{Hu:1994uz,Hu:1995en}. The evolution of four modes in the radiation dominated epoch is indicated in Fig.~\ref{fig: mixed-damping-mode} by pink, dashed lines.  Upon horizon crossing, neutrino perturbations receive the same boost from gravitational infall as dark matter. Being relativistic, neutrinos diffuse out of overdense regions and the competition between gravitational forces and the neutrino pressure leads to damped oscillations in $\delta_\upnu$ that can be clearly noticed in Fig.~\ref{fig: mixed-damping-mode}.

Photons and baryons are initially tightly coupled by Thomson scattering and oscillate as a single fluid. Upon recombination, the phase of the oscillation becomes imprinted on the CMB angular spectrum, causing the characteristic peak-though structure. Silk damping of the photon perturbations causes a decrease in height of the acoustic peaks at higher multipoles. The end of the baryon drag epoch is delayed with respect to photon decoupling \cite{Hu:1994uz,Hu:1995en}, and photon diffusion damps baryon fluctuations due to Compton drag. At the end of the Compton drag epoch, baryons fall into the dark matter potential wells, and fluctuations in the baryons become imprinted on the matter power spectrum as baryon acoustic oscillations (BAO).

While the introduction of neutrino-dark matter interactions could make the dynamics of this coupled sector to be equivalent to that of the baryon-photon plasma, an important difference arises due to the very different decoupling histories. The formation of neutral hydrogen at the epoch of recombination implies a steep decrease in the free electron fraction and hence in the photon baryon scattering rate. Still, decoupling of baryons and photons does not occur simultaneously but very closely. The Planck collaboration obtained $z_* = 1090\pm 0.41$ for the redshift of photon decoupling and $z_\mathrm{drag} = 1059.39\pm 0.46$ for the end of the baryon drag epoch \cite{Akrami:2018vks}. In contrast, the dark matter and the neutrino scattering rate evolve as power laws of the scale factor for the entire cosmological history relevant in this context, and the difference between the respective decoupling times can be considerable. Indeed, this is exactly the characteristic shaping  mixed damping - while neutrinos have already decoupled, dark matter is dragged along their free streaming evolution, and the growth of dark matter perturbations is inhibited.

\subsection{Analytical mixed damping evolution}
The basic premises of mixed damping allow for some simplifying assumptions which we use here to derive an analytical approximation for the evolution of dark matter perturbations. Firstly, mixed damping can only occur if dark matter decouples from neutrinos before matter radiation equality (c.f. Sec.~\ref{sec: parameter}). At these times, the metric potentials are dominated by radiation perturbations and are insensitive to alterations in the dark matter evolution to good approximation. Secondly, neutrinos evolve in the free-streaming regime and are coupled to all other components by gravitational interactions only. Hence, in the radiation dominated limit, neutrino perturbations are not affected by the modified dynamics of the dark matter sector. With these two assumptions, the dark matter evolution Eq.~(\ref{eq: dm-evolution}) can be reformulated in terms of an external source function $S\left(k\,,\tau\right)$
\begin{equation}
\ddot{\delta}_\dm + \left(\frac{1}{\tau} + \frac{C_\upkappa}{\tau^3}\right)\dot{\delta}_\dm = S\left(k\,,\tau\right)\,.
\label{eq: analytic-original-eq}
\end{equation}
During radiation domination, the reduced Hubble rate is $\ha=\tau^{-1}$ and for $n_{\nu\dm}=0$ we decompose the scattering rate as $R\,\dmu = C_\upkappa\,\tau^{-3}$, where $C_\upkappa$ is a constant. The source function depends on the metric potentials and on the neutrino velocity divergence as
\begin{equation}
S\left(k\,,\tau\right) = 3\ddot{\phi} + \frac{3}{\tau}\,\dot{\phi} - k^2\psi + \frac{C_\upkappa}{\tau^3}\left(3\dot{\phi} - \theta_\upnu\right)\,.
\label{eq: analytic-source}
\end{equation}
Neglecting the neutrino anisotropic stress implies $\phi=\psi$ (see Eq.~(\ref{eq: einstein})), and the metric perturbations deep in the radiation dominated epoch are approximated by \cite{Dodelson:2003ft}
\begin{equation}
\phi\left(k\,,\tau\right) = 3\,\phi_\mathrm{p}\left(k\right)\left(\frac{\sin\left(k\tau/\sqrt{3}\right) - \left(k\tau/\sqrt{3}\right)\cos\left(k\tau/\sqrt{3}\right)}{\left(k\tau/\sqrt{3}\right)^3}\right)\,,
\end{equation}
where $\phi_\mathrm{p}$ is the primordial magnitude of the fluctuation, i.e. $\phi\left(k,\tau\rightarrow0\right) = \phi_\mathrm{p}\left(k\right)$. For the neutrino free-streaming evolution we take \cite{Blas:2011rf}
\begin{equation}
\theta_\upnu\left(k\,,\tau\right) = \frac{3k}{4}\left.\left(\delta_\upnu + 4\psi\right)\right|_{\tau=0}\, j_1\left(k\tau\right)
+ 6k\int_0^\tau d\tau'\, \dot{\phi}\,j_1\left[k\left(\tau-\tau'\right)\right]\,.
\label{eq: analytic-neutrino-velocity}
\end{equation}
The full solution to the time evolution of dark matter perturbations (\ref{eq: analytic-original-eq}) is a linear combination of the homogeneous solutions and a particular solution, constructed from Green's method
\begin{equation}
\delta\left(k\,,\tau\right) = C_1 + C_2\, \mathrm{Ei}\left(\frac{C_\upkappa}{2\tau^2}\right)
+ \int_0^\tau d\tau' S\left(k\,,\tau'\right)\,\frac{\tau'^2}{2}\,\mathrm{e}^{-\frac{C_\upkappa}{2\tau'}}
\left[\mathrm{Ei}\left(\frac{C_\upkappa}{2\tau'^2}\right) - \mathrm{Ei}\left(\frac{C_\upkappa}{2\tau^2}\right)\right]\,.
\label{eq: delta-analytic}
\end{equation}
For small values of $\tau$, when the mode is well outside the Hubble radius, $\delta\left(k\,,\tau\right)$ is constant. Hence the initial conditions dictate $C_2=0$ and $C_1 = \delta\left(k\,,0\right)$. We evaluate the integral numerically for several modes at consecutive conformal times. The results are indicated as black crosses in Fig.~\ref{fig: mixed-damping-mode}. Notice that our final results capture all relevant qualitative features, even if our approximations are not perfect~\footnote{While the analytic form for the metric evolution follows the numerical results rather well, its first and second derivatives are less accurate. We also find that our expression for the neutrino velocity gets damped slightly slower than in the numerical case.}. Because our main intention in deriving the analytical solution is to shed light on the underlying physics, we keep  the simple analytic forms given above with no attempt to improve the accuracy using e.g. fitting formulas~\cite{Hu:1994uz,Hu:1995en}.

\subsection{Evolution of density perturbations and the matter power spectrum}
\label{susec: mixed-damping}

\begin{figure}
\centering
\includegraphics[]{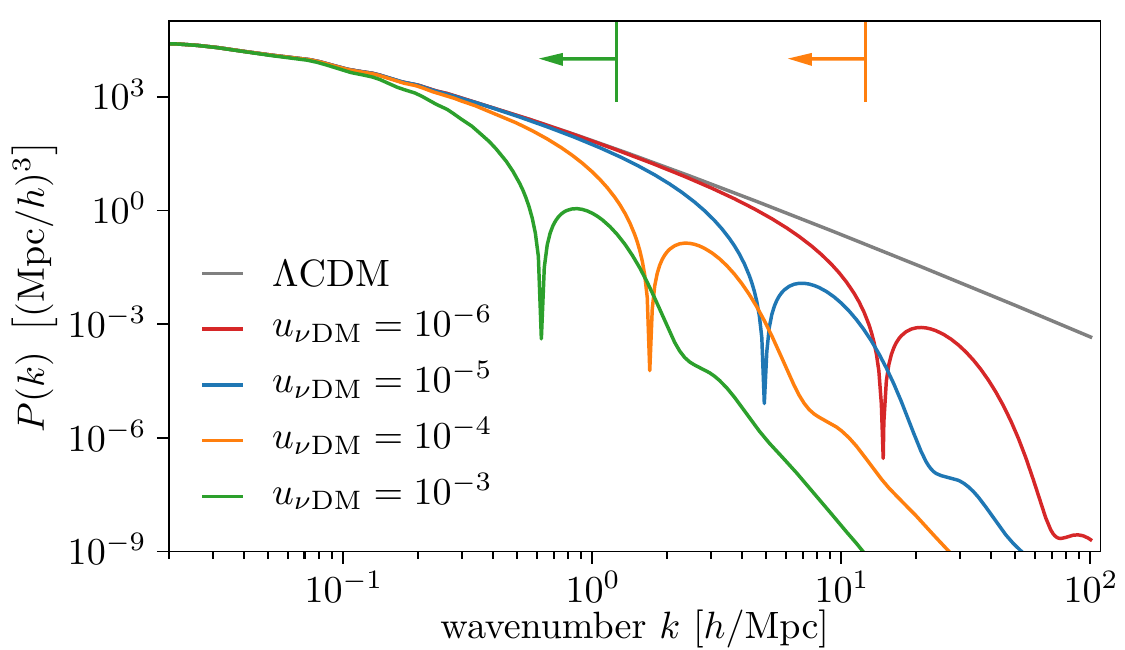}
\caption{The matter power spectrum in presence of dark matter-neutrino interactions. Depending on the size of a mode and the particular value of $\unudm$, the suppression with respect to $\Lambda$CDM is caused by mixed damping only or a by combination of collisional and mixed damping. Arrows in the same colours as the graphs indicate the wavenumbers for which mixed damping is the sole mechanism responsible for the suppression of power. For the smallest scattering rates shown here these bounds lie outside the plotted range.}
\label{fig: mixed-damping-pk}
\end{figure}

The matter power spectrum for several neutrino-dark matter interacting scenarios is shown in Fig.~\ref{fig: mixed-damping-pk}. For the largest values of $\unudm$, perturbations on small scales are subject to a mixture of collisional and mixed damping, while for larger wavenumbers and smaller scattering rates all power suppression with respect to $\Lambda$CDM is caused by mixed damping only. We indicate the transition between these regimes in Fig.~\ref{fig: mixed-damping-pk} by coloured arrows. In general, the matter power spectra show a common set of features, namely:

\begin{itemize}
	\item At the largest scales, or correspondingly for the smallest wavenumbers, the matter power spectrum does not differ from the $\Lambda$CDM case.
	\item Continuing to smaller wavenumbers, a steep decrease in power is followed by a bump after which the matter power spectrum continues to decrease again.
	\item A small plateau is encountered in the subsequent damping tail, after which the perturbations continue to decrease further.
\end{itemize}

The first of these features, i.e. the $\Lambda$CDM-like behaviour on large scales, can be easily understood. These scales enter the Hubble radius after dark matter has decoupled from the neutrinos and hence are not affected by the interaction. An estimate of the scale which enters the Hubble radius when $\Gamma_{\dm-\upnu}=H$ yields
\begin{equation}
k_{\dm,\mathrm{dec}} = 2\uppi\,H_0\,\sqrt{\Omega_\mathrm{r}} \simeq \frac{6.67\times 10^3}{\sqrt{\unudm}}\,h/\mathrm{Mpc}\,,
\end{equation}
for $\nnudm=0$, reproducing the onset of the damping rather well. In terms of the analytic solution of Eq.~(\ref{eq: delta-analytic}), it is reassuring to note that it recovers the $\Lambda$CDM behaviour in the limit $C_\upkappa \ll \tau^2$. In this limit the exponential integral can be expanded as
\begin{equation}
\mathrm{Ei}\left(\frac{C_\upkappa}{2\tau^2}\right) = \upgamma_\mathrm{E} + \ln\left(C_\upkappa/2\right) - 2\ln\left(\tau\right) + \mathcal{O}\left(\frac{C_\upkappa}{2\tau^2}\right)\,,
\label{eq: exponential-integral-expansion}
\end{equation}
and the first two terms can be absorbed in the definition of $C_1$. The source function $S\left(k\,,\tau\right)$, on the other hand, reduces to the $\Lambda$CDM case if the last term of Eq.~(\ref{eq: analytic-source}) can be neglected, such that one recovers the $\Lambda$CDM result (see e.g. Ref.~\cite{Dodelson:2003ft}).

We show several examples for the time evolution of neutrino and dark matter density perturbations at smaller wavenumbers for $\unudm=10^{-6}$ in Fig.~\ref{fig: mixed-damping-mode} and compare them to the $\Lambda$CDM evolution. The largest of the modes lies at a position in the matter power spectrum in which there is already a suppression of power but before the first bump. As previously discussed, gravitational infall upon horizon crossing causes a bump in both the neutrino and dark matter perturbations, and after this initial kick, the neutrino perturbation undergoes damped oscillations around the zero point. After horizon crossing in the radiation dominated era the gravitational potentials decay, hence the source function (\ref{eq: analytic-source}) becomes zero at late times and the dark matter evolution at $\tau > \tau_\mathrm{late}$ can be approximated as
\begin{equation}
\delta\left(k\,,\tau\right) = \delta(k,\tau_\mathrm{late}) + 2\,\left(\int_0^{\tau_\mathrm{late}} d\tau'\, S(k\,,\tau')\,\frac{\tau'}{2}\, \mathrm{e}^{-\frac{C_\upkappa}{2\tau'^2}}\right)\ln\left(\tau\right)\,,
\label{eq: analytic-decoupled}
\end{equation}
where we have used the decoupled limit, Eq.~(\ref{eq: exponential-integral-expansion}). The result indicates that, after dark matter decoupling in the radiation dominated era, the perturbation still grows $\propto \ln a$. The growth at late times is clearly noticeable in the top panel of Fig.~\ref{fig: mixed-damping-mode} for both the analytical and the numerical result. However, the proportionality constant in front of the logarithm is modified by the presence of the neutrino velocity in the source function, and also the value of $\delta(k,\tau_\mathrm{late})$ is altered in comparison to the $\Lambda$CDM result.

\begin{figure}
\centering
\includegraphics[]{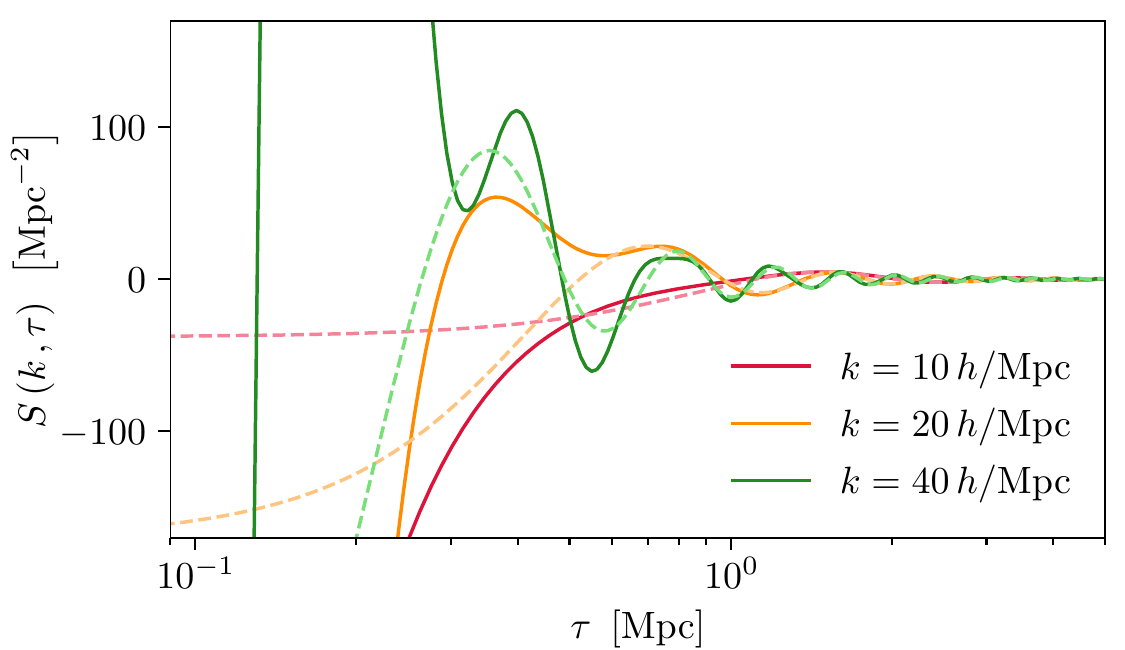}
\caption{External source function driving the evolution of dark matter perturbations at early times according to Eqs.~(\ref{eq: analytic-original-eq}) and (\ref{eq: analytic-source}). The two cases shown are $\Lambda$CDM (pastel, dashed lines) and an interacting scenario with $\unudm=10^{-6}$ (dark, solid lines).}
\label{fig: source}
\end{figure}

In the $\Lambda$CDM case, $\delta(k,\tau_\mathrm{late})$ and the integral in front of the logarithmic term in Eq.~(\ref{eq: analytic-decoupled}) have always negative values, therefore the dark matter density perturbations grow in the negative direction. This is evident for all modes in Fig.~\ref{fig: mixed-damping-pk}. The neutrino velocity modifies this behaviour, and its effect on the source function is larger as $k$ increases. Figure~\ref{fig: source} compares the source function for $\Lambda$CDM to that of an interacting model with $\unudm=10^{-6}$ for three values of $k$ illustrated previously in Fig.~\ref{fig: mixed-damping-mode} and clearly confirms this relation. The source function is modified more severely for larger values of $k$ due to the larger value of  $\theta_\upnu$, see Eq.~(\ref{eq: analytic-neutrino-velocity}). Notice from Fig.~\ref{fig: source} that a very steep decrease in the source function at early times appears in interacting scenarios. However, this gets largely suppressed by the smallness of the exponential function in the integral of Eq.~(\ref{eq: delta-analytic}). Instead, the relevant feature is the development of a positive peak which grows  with $k$ and can make the value of $\delta\left(k\,\tau_\mathrm{late}\right)$ positive. This is precisely what occurs for the second mode in Fig.~\ref{fig: mixed-damping-mode}, resulting from the fact that the dark matter fluid follows the neutrinos for long enough.

Finally, for the smallest mode of Fig.~\ref{fig: mixed-damping-mode}, the onset of logarithmic growth is delayed beyond the depicted time interval. Well after horizon crossing the gravitational potentials have decayed and oscillations in the neutrino density average to zero in the integral over the source function, i.e. the perturbation remains constant. An offset between $\delta_\upnu$ and $\delta_\dm$ can be noticed in Fig.~\ref{fig: mixed-damping-mode} for the smallest mode both in the analytical and the numerical evolution. This, however, does not indicate a decoupling between the two species but the fact that interactions are effective at the level of velocity perturbations, in such a way that $\theta_\dm$ closely tracks $\theta_\upnu$.

\section{Conclusions}
\label{sec: conclusions}

Non-standard dark matter scenarios, which predict a lower number of structures  at small-scales with respect to $\Lambda$CDM, have become popular in the last decade, as they have the potential to alleviate the various  problems that $\Lambda$CDM may face. This includes interacting dark matter scenarios with massless or very light Standard Model particles, such as photons or neutrinos. In these scenarios, dark matter fluctuations are usually erased by collisional damping. 

However,  there exists an additional (and equally relevant) damping regime, dubbed \textit{mixed damping}. Mixed damping is of particular relevance for dark mater-neutrino interaction scenarios. It takes place when dark matter remains coupled to neutrinos after they started to free-stream.  Extended Boltzmann codes, which describe interactions between dark matter and neutrinos, automatically account for the mixed damping effect. Nevertheless, the physical mechanism which leads to the suppression of structures in the mixed damping regime is distinct from the one operating in the collisional damping case. Our exploration of the parameter space shows that current constraints already exclude the parameter range for which only collisional damping is possible.

Future cosmological observations will be sensitive to very small scales and therefore will test even smaller dark mater-neutrino interaction rates. If these observations reveal that the matter power spectrum departs from $\Lambda$CDM due to neutrino-dark matter interactions, mixed damping is very important for the suppression of structure. In particular, larger modes, where the suppression sets in, are affected by mixed damping only, and this is where discrepancies are expected to be observed first. In contrast, smaller modes are subject to a mixture of collisional and mixed damping. Here, we provide a comprehensive study of the mixed damping effect. Comparing our analytic approximation for the evolution of dark matter density perturbations to the full numerical results, we can explain all features which appear in the damped matter power spectrum. Thus we are able to provide a clear physical picture of the mixed damping effect.

Beyond that, we also re-examine the numerical treatment of neutrino-dark matter interactions in the linear regime carefully. Our results reveal several discrepancies with previous studies. Higher-order terms in the Boltzmann hierarchy are not entirely independent from the specific form of the matrix element for the neutrino-dark matter interactions. We use a simplified model approach to characterise all possibilities and find that the angular dependence of the matrix elements does not influence the final spectrum of cosmological perturbations on a detectable level. We also revise the ultra-relativistic fluid approximation (UFA) in the presence of dark matter neutrino interactions and point out that it can only be consistently applied after neutrinos have decoupled from dark matter. If the UFA is not considered carefully, it causes unphysical artefacts at small scales in the matter power spectrum. Finally, we point out that the initial conditions for the numerical integration have to be revised in the presence of neutrino-dark matter interactions. We derive the appropriate expressions and test their effect on the CMB and matter power spectra. We find that neither the modifications to the UFA approach, nor the revised treatment of the initial conditions affects the theory predictions at a significant level. Hence, previous cosmological constraints on the strength of neutrino-dark matter interactions remain robust.

Last but not least, the interest and the reach of our results are not limited to the mixed damping regime. Rather, the correct description of the higher-order multipole coefficients, together with a suitable treatment of the UFA regime, are basic and indispensable pieces to analyse dark matter interactions with any light or massless degrees of freedom. The calculations carried out in this paper should be of broad interest for a large number of non-standard cosmological perturbation theory scenarios.

\acknowledgments
We thank Andrés Olivares-Del Campo for useful discussion and for sharing the matrix elements for neutrino-dark matter scattering in the simplified model approach. JS receives funding/support from the European Union’s Horizon 2020 research and innovation programme under the Marie Sklodowska-Curie grant agreement No 674896.OM is supported by the Spanish MINECO Grants FPA2017-85985-P and SEV-2014-0398. JS thanks IFIC and the Fermilab Theoretical Department for hospitality. This work is supported by the European Union’s Horizon 2020 research and innovation program under the Marie Sklodowska-Curie grant agreements No. 690575 and 674896.This research was supported in part by Perimeter Institute for Theoretical Physics. Research at Perimeter Instituteis supported by the Government of Canada through the Department of Innovation, Science, and EconomicDevelopment, and by the Province of Ontario through the Ministry of Research and Innovation.
\appendix

\bibliography{main}
\appendix
\section{Interaction formalism}
\label{sec:secA}
We present in Tab.~\ref{tab: interaction-scenarios} the values of $\dmu$ and $\alpha_l$ which allow for a proper calculation of high-order multipoles in the neutrino-dark matter interacting sector. We exploit the classification of dark matter-neutrino interaction scenarios provided in Ref.~\cite{Campo:2017nwh}.
\begin{sidewaystable}[ph!]
\centering
    \begin{tabular}{m{.15\textwidth} |m{.35\textwidth} |m{.15\textwidth}  |m{.2\textwidth} | m{.1\textwidth}}
        \makecell[c]{scenario}
        & 
        \makecell[c]{
        $\displaystyle \left(\frac{1}{\eta_\nu\, \eta_\dm}\sum_\mathrm{spins}\left|\mathcal{M}_{\nu\dm}\right|^2\right)$ }
        &
        \makecell[c]{
        $s = m_\dm^2 + 2m_\dm p_1$\\
        $t=2p_1^2(\mu-1)$\\
        $p_1 \ll m_M,\, m_\dm$
        }
        &
        \makecell[c]{scattering rate $\dmu$}
        &
        \makecell[c]{higher order\\ coefficients\\ $\alpha_l\,,~ l\ge2$}
        \\[0pt]\hline\hline
         
         \makecell{complex DM\\ Dirac mediator $N_R$}
         &
         \centering
         $\displaystyle
         g^4\, \frac{m_\dm^4 - 2sm_\dm^2+s^2+ts}{(t+s-m_N)^2}$
         &
         \centering
         $\displaystyle
         \frac{2g^4p_1^2m_\dm^2(\mu+1)}{(m_N^2 - m_\dm^2)^2}$
         &
         \centering
         $\displaystyle \frac{155}{588}\,
         \frac{a\,n_\dm\,\pi\,g^4}{(m_N^2 - m_\dm^2)^2}\,
         T_\nu^2$
         &
         \makecell{
         $\displaystyle \alpha_l = \frac{3}{2}$
         }
         \\[0pt] \hline
         
         \makecell[c]{real DM\\ Dirac mediator $N_R$}
         &
         \centering
         $\displaystyle
         g^4 \frac{\left(2s + t -2m_\dm^2\right)^2\left[m_\dm^4 - 2s m_\dm +s(s+t)\right]}
         {\left(s-m_N^2\right)^2\left(s + t +m_N^2 - 2m_\dm^2 \right)^2}$
         &
         \centering
         $\displaystyle
         \frac{32\,g^4\, m_\dm^4\, p_1^4\, \left(1+\mu\right)}
         {\left(m_N^2 - m_\dm^2 \right)^4}$
         &
         \centering
         $\displaystyle
         \frac{508}{21}\,\frac{g^4 \pi^3 m_\dm^2\, a\,n_\dm}
         {\left(m_N^2 - m_\dm^2\right)^4}\,
         T_\nu^4$
         &
         \makecell{
         $\displaystyle \alpha_l = \frac{3}{2}$
         }
         \\[0pt]\hline
         
         \makecell[c]{complex DM\\ Majorana mediator\\ $N_L$}
         &
         \centering
         $\displaystyle
         g^4\, \frac{m_\dm^4 - 2sm_\dm^2+s^2+ts}{(t+s-m_N)^2}$
         &
         \centering
         $\displaystyle
         \frac{2g^4p_1^2m_\dm^2(\mu+1)}{(m_N^2 - m_\dm^2)^2}$
         &
         \centering
         $\displaystyle \frac{155}{588}\,
         \frac{a\,n_\dm\,\pi\,g^4}{(m_N^2 - m_\dm^2)^2}\,
         T_\nu^2$
         &
         \makecell{
         $\displaystyle \alpha_l = \frac{3}{2}$
         }
         \\ \hline
         
         \makecell[c]{real DM\\ Majorana mediator\\ $N_L$}
         &
         \centering
         $\displaystyle
         g^4 \frac{\left(2s + t -2m_\dm^2\right)^2\left[m_\dm^4 - 2s m_\dm +s(s+t)\right]}
         {\left(s-m_N^2\right)^2\left(s + t +m_N^2 - 2m_\dm^2 \right)^2}$
         &
         \centering
         $\displaystyle
         \frac{32\,g^4\, m_\dm^4\, p_1^4\, \left(1+\mu\right)}
         {\left(m_N^2 - m_\dm^2 \right)^4}$
         &
         \centering
         $\displaystyle
         \frac{508}{21}\,\frac{g^4 \pi^3 m_\dm^2\, a\,n_\dm}
         {\left(m_N^2 - m_\dm^2\right)^4}\,
         T_\nu^4$
         &
         \makecell{
         $\displaystyle \alpha_l = \frac{3}{2}$
         }
         \\[0pt]\hline
         
         \makecell[c]{Dirac DM\\ scalar mediator $\phi$}
         &
         \centering
         $\displaystyle
         g^4\,\frac{\left(s + t - m_\dm^2\right)^2}
         {2\left(s + t + m_\phi^2 - 2m_\dm^2\right)}$
         &
         \centering
         $\displaystyle
         \frac{2\,g^4 m_\dm^2\, p_1^2}{\left(m_\dm^2 - m_\phi^2\right)^2}$
         &
         \centering
         $\displaystyle
         \frac{155}{392}\,
         \frac{\pi\,g^4\,a\,n_\dm}{\left(m_\phi^2 - m_\dm^2\right)^2}\,
         T_\nu^2$
         &
         \makecell{
         $\displaystyle \alpha_l = 1$
         }
         \\[0pt]\hline

         \makecell[c]{Majorana DM\\ scalar mediator $\phi$}
         &
         \centering
         $\displaystyle
         \frac{g^4}{2}\left[
         \frac{\left(s-m_\dm^2\right)^2}{\left(s-m_\phi^2\right)^2} +
         \frac{\left(s+t-m_\dm^2\right)^2}{\left(s+t+m_\phi^2-2m_\dm^2\right)^2}
         \right]$
         &
         \centering
         $\displaystyle
         \frac{4\,g^4\,m_\dm^2\,p_1^2}{\left(m_\phi^2-m_\dm^2\right)^2}$
         &
         \centering
         $\displaystyle
         \frac{155}{196}\,
         \frac{\pi\,g^4\,a\,n_\dm}{\left(m_\phi^2 - m_\dm^2\right)^2}\,
         T_\nu^2$
         &
         \makecell{
         $\displaystyle \alpha_l = 1$
         }
         \\[0pt]\hline
         
         \makecell[c]{vector DM\\ fermionic mediator\\ $N_L$}
         &
         \centering - - -
         &
         \centering
         $\displaystyle\frac
         {16\,g^4\, m_\dm^2\, p_1^2\, (3-\mu)}
         {3\left(m_\dm^2 - m_N^2\right)^2}$
         &
         \centering
         $\displaystyle
         \frac{1240}{441}\,
         \frac{\pi\,g^4\,a\,n_\dm}{\left(m_\dm^2 - m_N^2\right)^2}\,
         T_\nu^2$ 
         &
         \makecell{
         $\displaystyle \alpha_l = \frac{9}{8}$
         }
         \\[0pt]\hline

         \makecell[c]{scalar DM\\ vector mediator $Z_\mu$}
         &
         \centering
         $\displaystyle
         g^4\,\frac
         {4\left[m_\dm^2 - 2m_\dm^2 s + s(s+t)\right]}
         {\left(m_Z^2 - t\right)^2}$
         &
         \centering
         $\displaystyle\frac
         {8\,g^4\,m_\dm\, p_1^2\,(\mu+1)}
         {m_Z^4}$
         &
         \centering
         $\displaystyle
         \frac{3}{2}\, \frac{310}{441}\, \frac{\pi g^4\, a\, n_\dm}{m_Z^4}\,
         T_\nu^2$
         &
         \makecell{
         $\displaystyle \alpha_l = \frac{3}{2}$
         }
         \\[0pt]\hline
         
         \makecell[c]{Dirac DM\\ vector mediator $Z_\mu$}
         &
         \centering
         $\displaystyle
         g^4\,\frac{4\left(m_\dm^2-s\right)^2 + 4st + 2t^2}
         {\left(m_Z^2 - t\right)^2}$
         &
         \centering
         $\displaystyle
         8 g^4\frac{m_dm^2\,p_1^2\, (1+\mu)}{m_Z^4}$
         &
         \centering
         $\displaystyle
         \frac{155}{147}\,
         \frac{g^4\,\pi\,a\,n_\dm}{m_Z^4}\,
         T_\nu^2$
         &
         \makecell{
         $\displaystyle \alpha_l = \frac{3}{2}$
         }
         \\[0pt]\hline
         
         \makecell[c]{Majorana DM\\ vector mediator $Z_\mu$}
         &
         \centering
         $\displaystyle
         g^4 \frac{2m_\dm^4 + 2s^2 + 2st + t^2 - 4m_\dm^2(s+t)}
         {\left(t-m_z^2\right)^2}$
         &
         \centering
         $\displaystyle
         2g^4 \frac{m_dm^2\,p_1^2\,(3-\mu)}{m_Z^4}$
         &
         \centering
         $\displaystyle
         \frac{155}{441}\,
         \frac{\pi\,g^4\,a\,n_\dm}{m_z^4}\,
         T_\nu^2$
         &
         \makecell{
         $\displaystyle \alpha_l = \frac{9}{8}$
         }
         \\[0pt]\hline
         
         \makecell[c]{vector DM\\ vector mediator $Z_\mu$}
         &
         \centering
         - - -
         &
         \centering
         $\displaystyle
         48\,g^4\,\frac{m_\dm^2 p_1^2 (1+\mu)}{m_Z^4}
         $
         &
         \centering
         $\displaystyle
         \frac{310}{49}\,
         \frac{\pi\,g^4\,a\,n_\dm}{m_Z^4}\,
         T_\nu^2$
         &
         \makecell{
         $\displaystyle \alpha_l = \frac{9}{8}$
         }
         \\[0pt]\hline

    \end{tabular}\\
    \caption{Possible scenarios for dark matter interactions and the corresponding matrix element, scattering cross section and coefficients for the higher multipoles in the Boltzmann equations.}
    \label{tab: interaction-scenarios}
\end{sidewaystable}
\afterpage{\clearpage}

\section{Evaluation of the effect of the ultra-relativistic fluid approximation}
\label{sec: ufa-appendix}
With the ultra-relativistic fluid approximation the Boltzmann hierarchy for neutrino perturbations (Eqs.~(\ref{eq: boltzmann-nu})) is cut after the anisotropic shear once perturbations are well inside the horizon, i.e. when $k\tau > \left(k\tau\right)_\mathrm{UFA}$. The UFA truncation scheme in general differs from the ordinary truncation scheme for Boltzmann equations proposed in Ref.~\cite{Ma:1995ey}. Its implementation in CLASS reads \cite{Blas:2011rf}
\begin{equation}
\dot{\sigma}_\ur = -\frac{3}{\tau}\sigma_\ur + \frac{2}{3}\left(\theta_\ur - 6\dot{\phi}\right) \,.  \label{ufa: truncation-equation}
\end{equation}
where the subscript ``$\mathrm{ur}$'' refers to any ultra-relativistic species. Previous studies dealing with dark matter-neutrino interactions have tried to generalise the expression above as
\begin{equation}
\dot{\sigma}_\ur = -\frac{3}{\tau}\sigma_\ur + \frac{2}{3}\left(\theta_\ur - 6\dot{\phi}\right) - \dmu\sigma_\ur \,.
\label{ufa: truncation-equation2}
\end{equation}
However, we could not find a consistent derivation for the latter expression and adopt an alternative approach, in which we delay the onset of the UFA until neutrinos have decoupled. In the mixed damping scenario neutrinos have to decouple before matter radiation equality (note that $\acrit < a_\mathrm{eq}$). The scale factor at decoupling is defined by the condition $\Gamma_{\nu-\dm}(\anudec) \equiv H(\anudec)$ and approximately given by
\begin{equation}
a_{\upnu,\mathrm{dec}}^{\nnudm+1} =  \frac{3 \mp^2}{8\uppi}\frac{\Omega_\dm}{\sqrt{\Omega_\mathrm{r}}}\frac{\unudm\,\sigma_\mathrm{Th}\,H_0}{100\,\mathrm{GeV}}
= 1.19 \times 10^{-2}\times \unudmO \times\left(\frac{\Omega_\dm h^2}{0.1186}\right)\,.
\end{equation}
Correspondingly, the ultra-relativistic fluid approximation should not be employed before
\begin{equation}
\tau_\ufa = \frac{a_{\upnu,\mathrm{dec}}}{H_0\sqrt{\Omega_\mathrm{r}}}
= \begin{cases}
5.53 \times 10^3\,\mathrm{Mpc}\times \unudm \times\left(\frac{\Omega_\dm}{0.1186}\right) &\text{if } n_{\nu\dm} = 0\\
10.6 \times 10^4\,\mathrm{Mpc}\times \unudm^\frac{1}{3}\times \left(\frac{\Omega_\dm}{0.1186}\right)^\frac{1}{3} &\text{if } n_{\nu\dm} = 2\\
19.2 \times 10^4\,\mathrm{Mpc}\times \unudm^\frac{1}{5}\times\left(\frac{\Omega_\dm}{0.1186}\right)^\frac{1}{5} &\text{if } n_{\nu\dm} = 4
\end{cases}\,,
\end{equation}
The trigger value $\left(k\tau\right)_\mathrm{UFA}$ for the onset of the UFA is determined by the wavenumber $k_\mathrm{max}$ of the smallest mode we are interested in, and we require
\begin{equation}
k_\mathrm{max}\,\tau_\mathrm{UFA} \ge \left(k\tau\right)_\mathrm{UFA}\,.
\end{equation}
Before the onset of the UFA, the Boltzmann hierarchy is cut at some large multipole $l_\mathrm{max}$. To avoid unphysical reflections in this regime we make sure to choose $\lm \ge \left(k \tau\right)_\mathrm{UFA}$.

We consider two benchmark scenarios to investigate the impact of the UFA approach on the CMB temperature auto-correlation, E-mode polarisation auto-correlation and the temperature-E-mode cross-correlation spectra. Namely, the parameters we consider are $\nnudm=0$ and  $\unudm=4.5\times 10^{-5}$ (upper limit from Ref.~\cite{DiValentino:2017oaw}) and $\nnudm=2$ and $\unudmO=5.4\times10^{-14}$ (upper limit derived from  Ref.~\cite{Wilkinson:2014ksa}).  The remaining six $\Lambda$CDM parameters are set to the mean values obtained in the Planck 2018 data release \cite{Aghanim:2018eyx}, and we consider three massless neutrinos with identical interactions to dark matter. For each scenario, we delay the UFA regime by increasing the value of $\left(k\tau\right)_{\mathrm{UFA}}$, both varying and keeping fixed the value of $\lm$. We find that the effect of delaying the UFA regime on the CMB spectra is completely negligible for both scenarios.

\begin{figure}
	\centering
	\includegraphics{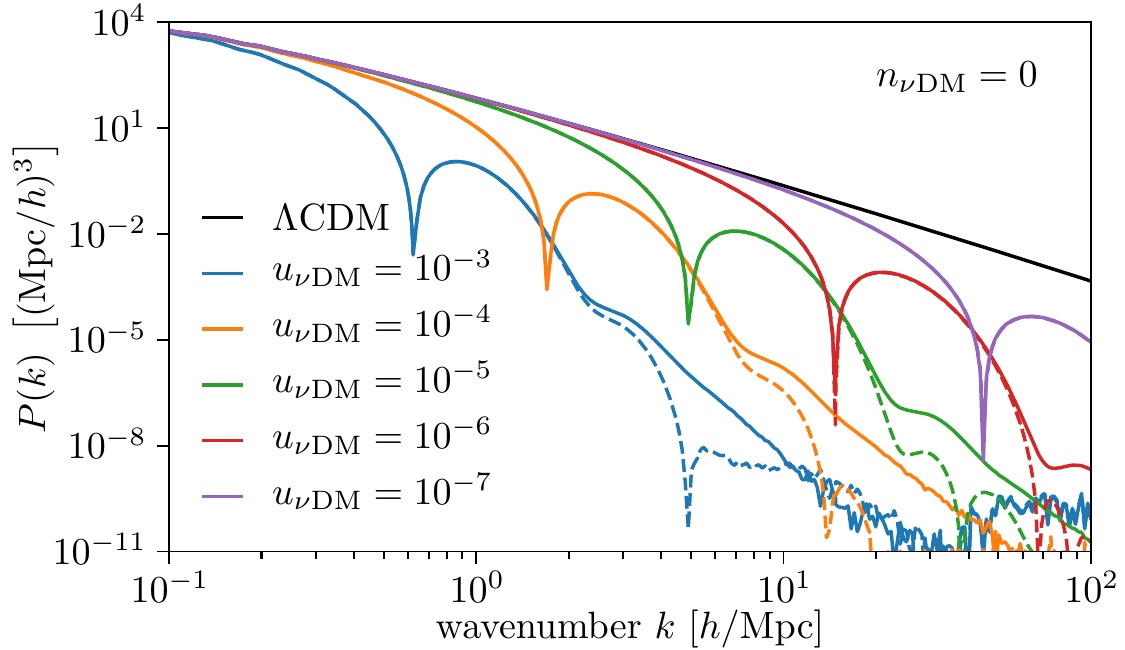}
	\caption{The matter power spectrum computed with the default CLASS UFA settings (dashed lines) and with a delayed onset of the UF approximation ($k\tau= \lm= 200$, solid lines). Non-negligible differences appear beyond the first oscillation peak.}
	\label{fig: ufa-impact-pk}
\end{figure}

However, the impact on the matter power spectrum is clearly noticeable, as we shall now illustrate, focusing exclusively on the $\nnudm=0$ case. Figure~\ref{fig: ufa-impact-pk} shows that, up to the first oscillation peak, the power spectra computed with CLASS UFA default settings, (Eq.~(\ref{ufa: truncation-equation}), see dashed lines) and with a delayed UFA regime approach (solid lines) typically agree. However, at smaller scales, non-negligible discrepancies show up. We also compare the results obtained with a truncation according to Eq.~(\ref{ufa: truncation-equation}) and Eq.~(\ref{ufa: truncation-equation2}) for the default and the delayed UFA settings. Note that Ref.~\cite{Wilkinson:2014ksa} uses the truncation scheme of Eq.~(\ref{ufa: truncation-equation2}) but with default CLASS parameters for the onset of the UFA. Either truncation scheme results in a very similar power spectrum if the UFA regime is delayed sufficiently. This behaviour is expected since, in this case, the scattering term in Eq.~(\ref{ufa: truncation-equation2}) is small and should have no impact. On the other hand, increasing $\lm$ while not delaying the onset of the UFA method leaves the power spectrum unchanged. Most importantly, however, the modified truncation scheme of Ref.~\cite{Wilkinson:2014ksa} is not able to reproduce the result to which both codes converge for a delayed UFA regime for default UFA parameters on the smallest scales. We therefore conclude that, to obtain precise predictions for the matter power spectrum on small scales, it is indeed necessary to delay the onset of the UFA regime. The more economical approach of including the interaction term to the truncation equation (\ref{ufa: truncation-equation2}) can not reproduce the correct small scale behaviour.

As the power law of the dark matter-neutrino interaction cross section steepens, i.e. $\nnudm$ in Eq.~(\ref{eq: nnudm}) increases, the duration of the mixed damping regime is shortened. This trend is clearly visible in Fig.~\ref{fig: parameterspace-scattering-rates} (see also Eq.~(\ref{eq: nu-decoupling-n-dependence}) and the discussion there). Hence, the time interval during which default UFA setting would erroneously treat neutrinos as free streaming particles tends to be shorter for larger values of $\nnudm$. Nevertheless, we advocate to conduct a careful convergence study, as the accuracy of the results will depend on the precise combination of $\unudmO$ and $\nnudm$ considered.

Even though the default UFA settings fail to correctly predict the matter power spectrum on small scales, we do not expect that these discrepancies affect current or future constraints from observations of the power spectrum. Differences at this level will likely be erased by non-linear effects during structure formation and, in addition, they are well below the expected statistical and systematic uncertainties. To further illustrate this point, we use the matter power spectrum with both the default and the delayed UFA settings to estimate the number of Milky Way satellites following Ref.~\cite{Gariazzo:2017pzb}
	~\footnote{See also the works of Refs.~\cite{Boehm:2014vja,Schewtschenko:2014fca,Schewtschenko:2015rno} for devoted simulations within different possible interacting dark matter scenarios.}.
Table~\ref{tab: ufa-nsat} summarises the results in the  $\nnudm=0$ case for different values of the neutrino-dark matter interaction parameter $\unudm$. Differences at this level are negligible compared to the expected Poisson scatter, systematic uncertainties from baryonic feedback and/or the modelling of non-linear structure formation.

\begin{table}
	\centering
	\renewcommand{\arraystretch}{1.2}
	\begin{tabular}{c|c|c}
		scenario  &  default UFA settings & delayed UFA regime \\
		\hline
		$\Lambda$CDM  &  160 & 160 \\
		$\unudm=1\times10^{-7}$ & 30.8 & 30.7 \\
		$\unudm=5\times10^{-7}$ & 8.5 &  8.5 \\
		$\unudm=1\times10^{-6}$ & 5.8 & 5.9 \\
		$\unudm=5\times10^{-6}$ & 0.72 & 0.74\\
		$\unudm=5\times10^{-5}$ & 0.05 & 0.06 \\
	\end{tabular}
	\caption{Number of Milky Way satellite galaxies $N_\mathrm{sat}$ computed from the linear matter power spectra obtained with default UFA settings and with a delayed UFA regime ($k\tau=\lm=200$). We consider the $\nnudm=0$ case for different dark matter-neutrino cross sections.} 
	\label{tab: ufa-nsat}
\end{table}

\section{Initial conditions for coupled neutrinos}
\label{sec: ini-appendix}
The initial conditions for cosmological perturbations in CLASS are given to second order in $k\tau$ and zeroth order in the inverse Thomson scattering rate \cite{CyrRacine:2010bk}. These expressions need to be modified when neutrinos interact with dark matter, because the scattering suppresses the neutrino anistotropic stress. To derive the modified expressions we follow the steps of \cite{Ma:1995ey} and obtain
\begin{subequations}
\begin{align}
\phi_\mathrm{ini} = \psi_\mathrm{ini} = \frac{4}{3}\,C_\mathrm{ini}
\,,\\
\theta_{\upgamma,\mathrm{ini}} = \delta_{\upnu,\mathrm{ini}} = \frac{4}{3}\,\delta_{\mathrm{b},\mathrm{ini}} = \frac{4}{3}\,\delta_{\dm,\mathrm{ini}} = -\frac{8}{3}\,C_\mathrm{ini}\,,
\,,\\
\theta_{\upnu,\mathrm{ini}} = \theta_{\gamma,\mathrm{ini}} = \theta_{\dm,\mathrm{ini}} = \theta_{\mathrm{b},\mathrm{ini}} = -\frac{C_\mathrm{ini}}{18}\,k^4\tau^3
\,,\\
\sigma_{\upgamma,\mathrm{ini}} = \sigma_{\upnu,\mathrm{ini}} = 0\,,
\end{align}
\label{eq: ini-coupled-nu}
\end{subequations}
where the subscript ``b'' refers to baryons and $C_\mathrm{ini}$ is an integration constant.

\begin{figure}
\centering
\includegraphics{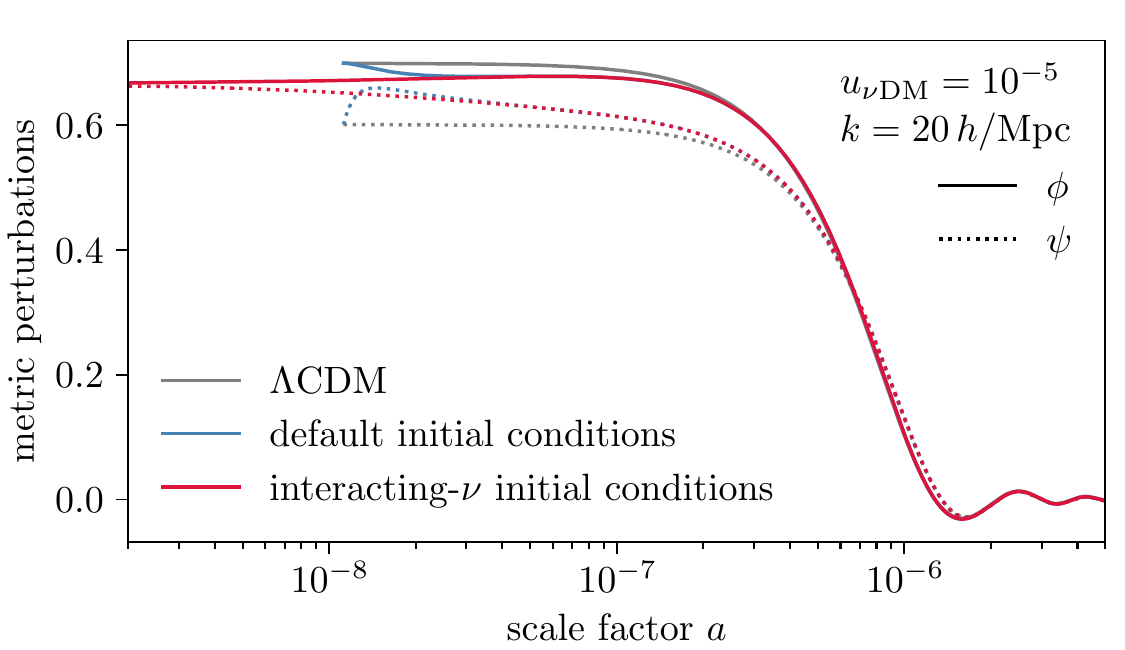}
\caption{
Comparison of the evolution of metric perturbation with the default and the revised initial conditions. For the case of revised initial conditions the integration starts at an earlier time to ensure that the approximation of neutrino-dark matter tight coupling is still valid.}
\label{fig: metric-initial-conditions}
\end{figure}

To further illustrate how dark matter-neutrino scattering affects the initial conditions of the metric perturbations we show their evolution for a single mode in Fig.~\ref{fig: metric-initial-conditions}. The mode has a wavenumber of $k=20\,h/\mathrm{Mpc}$ and enters the horizon during radiation domination, close to the time when neutrinos decouple. The numerical solution starts earlier if the initial conditions take into account neutrino interactions because Eq.~(\ref{eq: ini-coupled-nu}) is only applicable in the limit of tight coupling between dark matter and neutrinos. Notice from Fig.~\ref{fig: metric-initial-conditions} that the two treatments of the initial conditions converge to a common evolution quickly and therefore in our final result we observe that the choice of initial conditions has a negligible impact.

\end{document}